\documentclass[12pt,a4paper,keywords,prb]{revtex4-1}
\usepackage{subfigure}
\usepackage{graphicx}
\usepackage{amsmath}
\usepackage{epstopdf}
\usepackage{amsbsy}
\usepackage{color}
\usepackage{dcolumn}% Align table columns on decimal point
\usepackage{bm}% bold math

\begin{document}

\title{Local magnetization nucleated by non-magnetic impurities in Fe-based superconductors}
\author{Maria N. Gastiasoro and Brian M. Andersen}
\affiliation{Niels Bohr Institute, University of Copenhagen, Universitetsparken 5, DK-2100 Copenhagen,
Denmark}

\begin{abstract}

We study impurity-induced magnetic order within a five-band Hubbard model relevant to the normal paramagnetic phase of iron-based superconductors. The existence of the local magnetic order is explained in terms of an impurity-enhancement of states near the Fermi level, and we map out the resulting phase diagram of the existence of magnetization as a function of impurity strength and Coulomb correlations. In particular, the presence of impurity-induced magnetism in only a certain range of potential scattering strengths can be understood from the specific behavior of the impurity resonant state.

\end{abstract}

\date{\today}
\maketitle

\section{Introduction}

In a series of recent papers we have studied the consequences of local impurity-induced magnetic order in Fe-based systems. Specifically this concerns 1) the effect of the induced magnetization on the local density of states near potential scatterers\cite{Mng1,Mng2}, 2) unusual RKKY couplings of magnetic impurities\cite{MngMn}, 3) the emergence of nematogen impurities in the spin-density wave (SDW) phase\cite{Mng3},  and 4) the generation of elongated scattering centers in the nematic phase of the pnictides\cite{Mng4}. The latter two properties are directly related to the existence of electronic C$_2$ dimer states in underdoped pnictides\cite{chuang10,zhou11,song11,grothe12,hanaguri12,song12,allan13,rosenthal13}, and has been recently proposed as a crucial player in explaining e.g. the temperature dependence of the resistivity anisotropy of the 122 systems.\cite{tanatar,chu,chu,ishida13,Mng4,Mng5}  

The slowing down of magnetic fluctuations and subsequent induction of static short-range magnetic order near spatial inhomogeneities has been extensively discussed within one-band models relevant for cuprates.\cite{alloul09} This applies to non-magnetic disorder\cite{Tsuchiura,Wang,zhu02,Chen,harter07,andersen07,andersen10}, grain boundaries\cite{Tricoli}, and vortices\cite{Arovas,Andersen00,Chen02,Zhu,Takigawa,Andersen09}. Typically, in these cases $d$-wave superconductivity is crucial for the generation of local magnetic order due to the significant enhancement of the local density of states (LDOS) near the Fermi level only in the presence of the superconducting gap allowing for disorder-induced bound or resonant states. This is not the case for the typical models applicable to iron pnictides as will be discussed below.

Here, we revisit the properties of local magnetic order induced by non-magnetic point-like impurities in the iron pnictides. We obtain the general phase diagram for induced order, and explain it in terms of a strongly local modified density of states near the Fermi level by the impurities.

\section{Model}

The five-orbital model Hamiltonian is given by
\begin{equation}
 \label{eq:H}
\mathcal{H}=\mathcal{H}_{0}+\mathcal{H}_{int}+\mathcal{H}_{imp}.
\end{equation}
The first term is a tight-binding model,
\begin{equation}
 \label{eq:H0}
\mathcal{H}_{0}=\sum_{\mathbf{ij},\mu\nu,\sigma}t_{\mathbf{ij}}^{\mu\nu}c_{\mathbf{i}\mu\sigma}^{\dagger}c_{\mathbf{j}\nu\sigma}-\mu_{0}\sum_{\mathbf{i}\mu\sigma}n_{\mathbf{i}\mu\sigma}.
\end{equation}
Here the operators $c_{\mathbf{i} \mu\sigma}^{\dagger}$ ($c_{\mathbf{i}\mu\sigma}$) create (annihilate) an electron at the $i$-th site in the orbital $\mu$ and with spin projection $\sigma$, and $\mu_{0}$ is the chemical potential.
The indices $\mu$ and $\nu$ run through 1 to 5 corresponding to the five $d_{xz}$, $d_{yz}$, $d_{x^2-y^2}$, $d_{xy}$ and $d_{3z^2-r^2}$ iron 3d orbitals. 
The hopping integrals $t_{\mathbf{ij}}^{\mu\nu}$ are the same as those in  Ikeda \emph{et al.}~\cite{ikeda10}, included up to fifth nearest neighbors.

The third term in Eq.(\ref{eq:H}) describes the onsite Hubbard interaction
\begin{align}
 \label{eq:Hint}
 \mathcal{H}_{int}&=U\sum_{\mathbf{i},\mu}n_{\mathbf{i}\mu\uparrow}n_{\mathbf{i}\mu\downarrow}+(U'-\frac{J}{2})\sum_{\mathbf{i},\mu<\nu,\sigma\sigma'}n_{\mathbf{i}\mu\sigma}n_{\mathbf{i}\nu\sigma'}\\\nonumber
&\quad-2J\sum_{\mathbf{i},\mu<\nu}\vec{S}_{\mathbf{i}\mu}\cdot\vec{S}_{\mathbf{i}\nu}+J'\sum_{\mathbf{i},\mu<\nu,\sigma}c_{\mathbf{i}\mu\sigma}^{\dagger}c_{\mathbf{i}\mu\bar{\sigma}}^{\dagger}c_{\mathbf{i}\nu\bar{\sigma}}c_{\mathbf{i}\nu\sigma},
\end{align}
including  the intraorbital (interorbital) on-site repulsion $U$ ($U'$), the Hund's coupling $J$ and the pair hopping energy $J'$. We assume spin rotational invariance $U'=U-2J$, $J'=J$, and fix $J=U/4$.
Finally, $\mathcal{H}_{imp}=V_{imp}\sum_{\mu\sigma}c_{\mathbf{i^*}\mu\sigma}^{\dagger}c_{\mathbf{i^*}\mu\sigma}$ is the impurity potential, adding a potential energy $V_{imp}$ at the impurity site $\mathbf{i^*}$. We neglect the orbital dependence of the impurity potential for simplicity.
After mean-field decoupling of Eq.~\eqref{eq:Hint}, we solve the following eigenvalue problem $\sum_{\mathbf{j}\nu}
H^{\mu\nu}_{\mathbf{i} \mathbf{j} \sigma}
 u_{\mathbf{j}\nu\sigma}^{n}
=E_{n\sigma} u_{\mathbf{i}\mu\sigma}^{n}$,
where
\begin{align}
H^{\mu\nu}_{\mathbf{i} \mathbf{j} \sigma}&=t_{\mathbf{ij}}^{\mu\nu}+\delta_{\mathbf{ij}}\delta_{\mu\nu}[-\mu_0+\delta_{\mathbf{ii^*}}V_{imp}+U \langle n_{\mathbf{i}\mu\bar{\sigma}}\rangle\nonumber\\
&+\sum_{\mu' \neq \mu}(U'\langle n_{\mathbf{i}\mu' \bar{\sigma}}\rangle+(U'-J)\langle n_{\mathbf{i}\mu' \sigma}\rangle)],
 \end{align}
on a $20\times 20$ lattice with self-consistently obtained densities $\langle n_{\mathbf{i}\mu\sigma} \rangle=\sum_{n}|u_{\mathbf{i}\mu\sigma}^{n}|^{2}f(E_{n\sigma})$ for each site and orbital.

\begin{figure}[t]
 \includegraphics[clip=true,width=0.7\columnwidth]{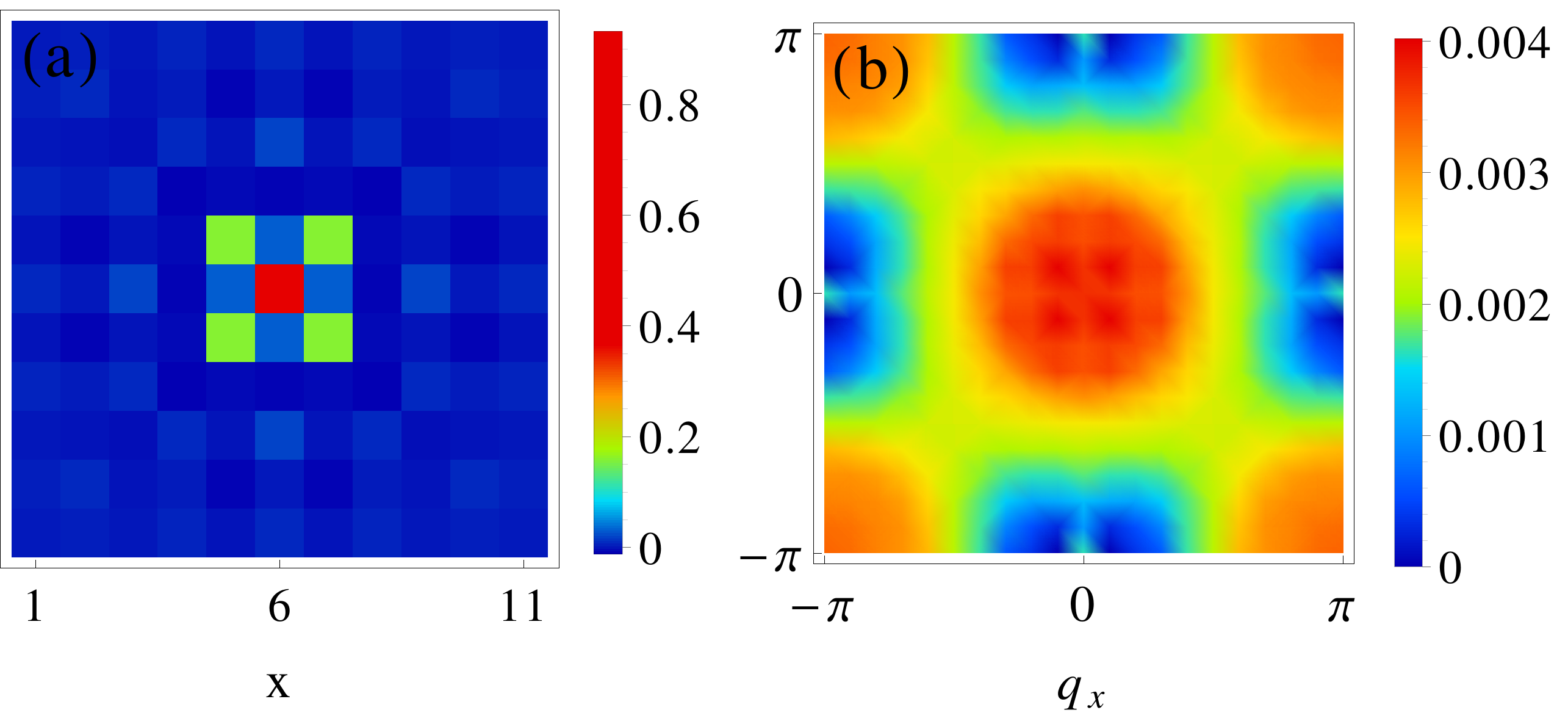}
\caption{(a) Induced real-space magnetization around the impurity, and (b) corresponding momentum space representation, for $U<U_{c2}=0.8$ eV and $V_{imp}=1.3$ eV.} 
\label{fig:1}
\end{figure}

\section{Results and Discussion}

The mean field ground state of the Hamiltonian~\eqref{eq:H} without disorder can be controlled by the interaction parameter $U$. 
For $U$ lower than a critical value of $U_{c2} \sim 0.9$ eV, the undoped homogeneous system remains non-magnetic. Above $U_{c2}$ the system is unstable to $(\pi,0)$ [or $(0,\pi)$] ordering. Depending on parameters it can, however, be energetically favorable to spin polarize regions around the non-magnetic impurities even in the normal paramagnetic state. Figure~\ref{fig:1}(a) shows the real space magnetization in a case where impurity-induced order is present. In this particular case, the impurity site itself exhibits the largest induced moment in contrast to the similar one-band result where the nearest-neighbor sites contain the largest spin polarization\cite{Tsuchiura,Wang,zhu02,Chen,harter07,andersen07,andersen10}. The Fourier transform of the magnetization in Fig.~\ref{fig:1}(a) is shown in Fig.~\ref{fig:1}(b). 
It is dominated by broad peaks at $(0,0)$ and $(\pi,\pi)$, and weaker but sharp peaks at $(0,\pi)/(\pi,0)$. The latter originate from the longer-range but weaker tails of the induced magnetization\cite{MngMn}.

\begin{figure}[t]
\includegraphics[clip=true,width=0.6\columnwidth]{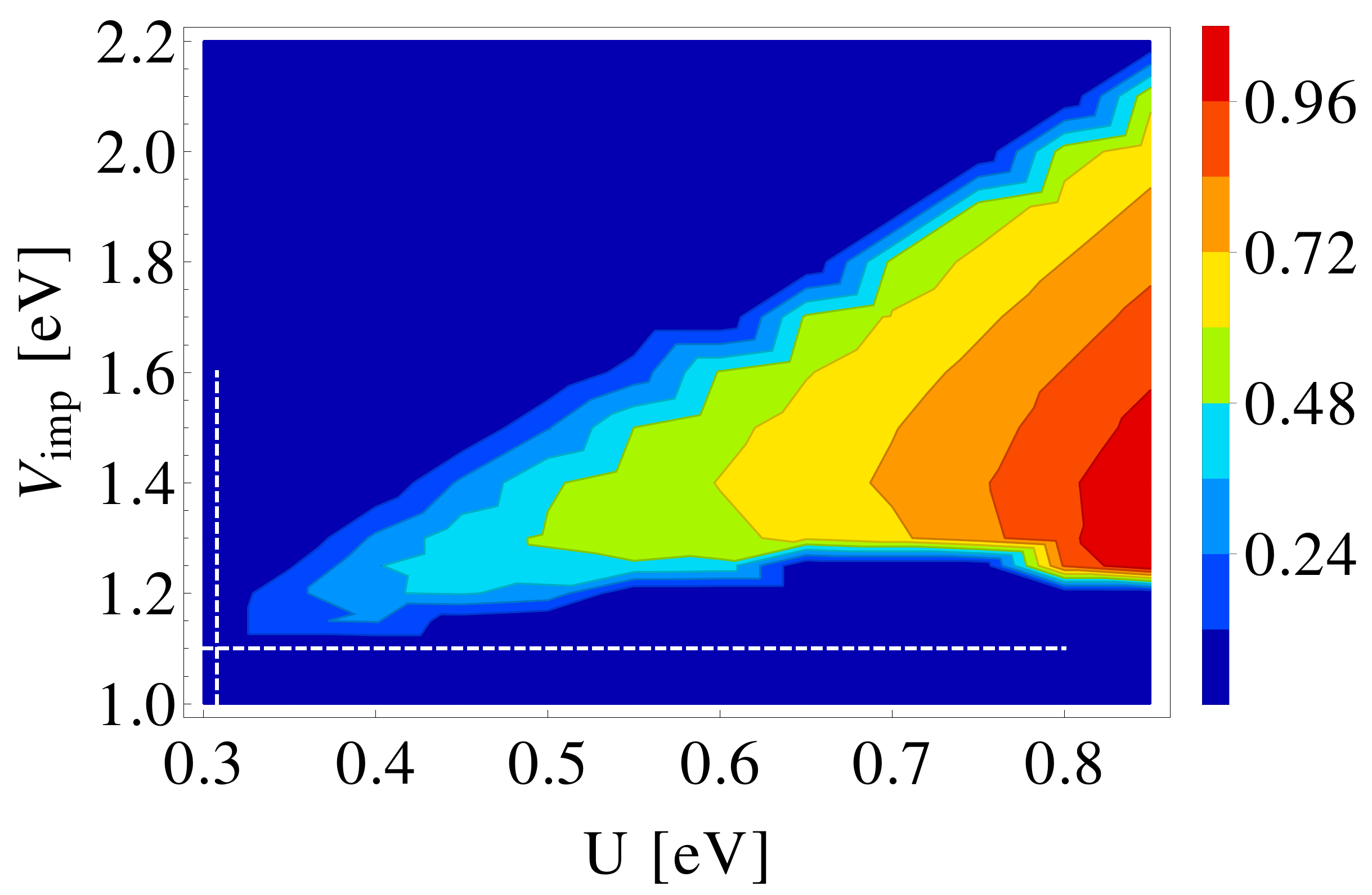}
\caption{Phase diagram of impurity-induced magnetization at the impurity site as a function of $U$ and $V_{imp}$ in the  normal state.} 
\label{fig:2}
\end{figure}

In order to map out when magnetization is locally nucleated, we explore the parameter space by varying the interaction $U$ and impurity strength $V_{imp}$. For the band structure of Ref. \onlinecite{ikeda10}, only repulsive potentials, $V_{imp}>0$, are capable of inducing magnetic order locally at zero temperature and $n=6.0$ filling.
The resulting moment at the potential site is shown in Fig.~\ref{fig:2}. The triangular-shaped region of finite impurity-induced magnetization exhibits a clear asymmetry by a linear (curved) phase boundary in the large (low) $V_{imp}$ limit. It is noteworthy that due to the curved lower phase boundary, there exist values of the potential $V_{imp}$ (around $\sim 1.2$ eV for the present band structure and doping level) where it is favorable to locally nucleate magnetism only for weak values of $U$, contrary to naive expectations from a standard Stoner instability criterion.

\begin{figure}[t]
 \includegraphics[clip=true,width=0.55\columnwidth]{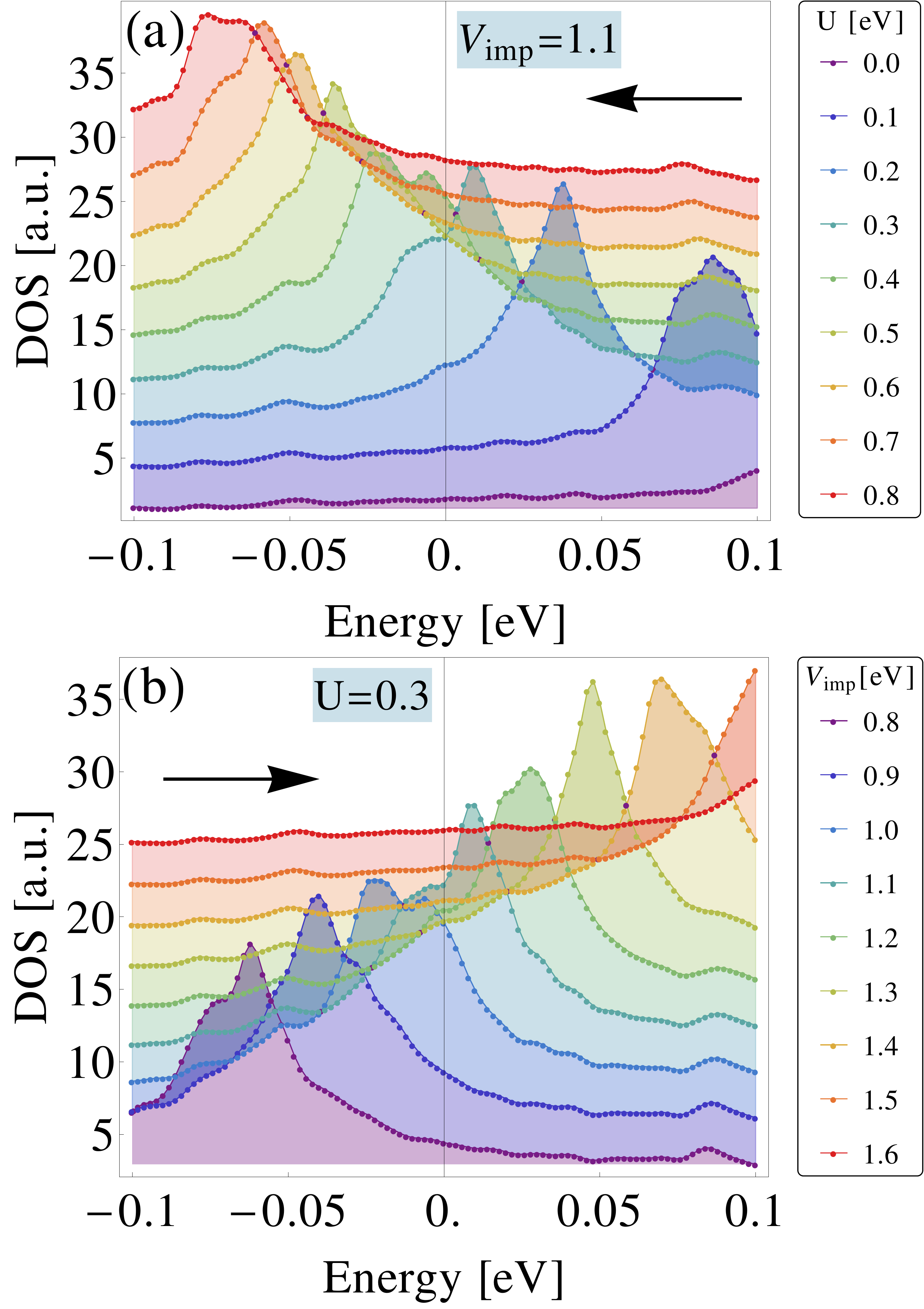}
\caption{LDOS versus energy at the impurity site for (a) $V_{imp}=1.1$eV and several $U$, and (b) for $U=0.3$eV and a representative range of potential strengths $V_{imp}$ as shown in the legend.} 
\label{fig:3}
\end{figure}

The above results can be understood by studying the evolution of the LDOS in the presence of the impurity. 
Specifically, by tracking the formation of impurity resonant states close to the Fermi energy ($\epsilon_F$) as a function of $U$ and $V_{imp}$ at the impurity site. 
We focus initially on the effect of correlations $U$ for a fixed potential $V_{imp}=1.1$ eV (horizontal dashed white line in Fig.~\ref{fig:2}) with results shown in Fig.~\ref{fig:3}(a). 
Without correlations ($U=0$), the resonant state exists at high positive energies ($>0.1$ eV). 
For increasing $U$, however, as seen from Fig.~\ref{fig:3}(a) the resonant state is pushed to lower energies, crosses $\epsilon_F$, and moves to negative energies. The largest density of states near $\epsilon_F$ is found around $U=0.3$ eV which explains the curved lower edge of the magnetization region.

Figure~\ref{fig:3}(b) shows the LDOS at the impurity site for $U\sim 0.3$ eV for different values of the impurity strength $V_{imp}$ (vertical dashed white line in Fig.~\ref{fig:2}). 
Evidently, the resonant state exhibits the opposite trend to Fig.~\ref{fig:3}(a) by moving to higher energies upon increasing the impurity strength. Thus, the enhanced LDOS at $\epsilon_F$ by the resonant state is maximal inside the triangular wedge-shaped region of Fig.~\ref{fig:2}, and the fact that impurity-induced magnetization exists in this same region agrees well with the Stoner instability criterion.

The reason resonant states can exist in the normal state is due to the "gapped" $e_g$ orbitals near the Fermi level. 
For the present band, the density of states of e.g. the $d_{3z^2-r^2}$ orbital is negligible in the energy range of Fig.~\ref{fig:3}, and hence it is possible to generate reasonably sharp resonant states of mainly $d_{3z^2-r^2}$ character in this energy window. 
We have verified that indeed the orbital content of the resonant states of Fig.~\ref{fig:3} (as well as the onsite magnetization) is almost entirely of $d_{3z^2-r^2}$ character. 
This is unlike the case of e.g. LiFeAs where the induced magnetization exists mainly on nearest-neighbor sites, and is made up from the $d_{xz}$ and $d_{yz}$ orbitals\cite{Mng1}. 
These results highlight the importance of using realistic five-band models in studies of impurity-induced local order of the iron pnictides.

\section{Conclusions}

In summary, we have discussed the generation of local magnetic order in the normal state of the Fe-based superconductors within a five-band model that includes onsite Coulomb interactions at the mean-field level. We have mapped out the phase diagram as a function of $U$ and $V_{imp}$ for the existence of local magnetic order, and explained the main topology of the phase diagram in terms of impurity-resonant states near the Fermi level.

\section{Acknowledgement}

We acknowledge support from the Lundbeckfond fellowship (grant A9318).

\end{document}